\documentclass[12pt]{article}
\usepackage{jheppub}
\usepackage{bbm}
\usepackage{epsfig, amsmath, amssymb, dsfont, mathtools}
\usepackage{hyperref}

\hypersetup{
    colorlinks=true,
    linkcolor=blue,
    citecolor=blue,
    urlcolor=blue
}
\usepackage{graphicx,psfrag}
\usepackage{xcolor}
\usepackage{physics}
\usepackage[utf8]{inputenc}
\newcommand{\nn}{\nonumber}
\newcommand{\be}{\begin{equation}}
\newcommand{\ee}{\end{equation}}
\newcommand{\ben}{\begin{equation}}
\newcommand{\een}{\end{equation}}
\newcommand{\bea}{\begin{eqnarray}}
\newcommand{\eea}{\end{eqnarray}}
\newcommand{\bA}{\begin{array}}
\newcommand{\eA}{\end{array}}
\newcommand{\bc}{\begin{center}}
\newcommand{\ec}{\end{center}}
\newcommand{\al}{\alpha}

\newcommand{\ra}{\rightarrow}

\newcommand{\ie}{{\it i.e.}}
\newcommand{\eg}{{\it e.g.}}

\newcommand{\lan}{\langle}
\newcommand{\ran}{\rangle}

\newcommand{\scri}{{\mathcal I}}

\begin{document}


\begin{titlepage}

%

\bc

\hfill 
\\         [15mm]

{\Huge de Sitter extremal surfaces, time contours, 
   \\ [2mm]
   complexifications and pseudo-entropies} \\
\vspace{12mm}
{\large K.~Narayan} \\
\vspace{3mm}
{\small \it Chennai Mathematical Institute, \\}
{\small \it H1 SIPCOT IT Park, Siruseri 603103, India.\\}

\ec
\vspace{30mm}

\begin{abstract}
  ~We study no-boundary de Sitter extremal surfaces and their
  pseudo-entropy areas for generic subregions at the future boundary,
  building on previous work. For large subregions, timelike+Euclidean
  extremal surfaces exist with transparent geometric interpretations,
  as do complex ones. The situation for small subregions is analogous
  to Poincare $dS$ and only complex extremal surfaces exist.  In
  general, the extremal surface area integrals are defined via time
  contours in the complex time plane. We find multiple extremal
  surfaces with indistinguishable areas whose time contours can be
  deformed into each other in the complex time plane without
  obstruction, which are equivalent for these purposes. This also
  suggests equivalences between complex $dS$ replica geometries. We
  discuss $dS_3$ as a simple example at length. This suggests a
  picture for multiple subregions and entropy inequalities in de
  Sitter, as encoding $AdS$ ones via analytic continuation. We also
  discuss mapping future boundary subregions and those on constant
  time slices in the static patch via lightrays.
\end{abstract}


\end{titlepage}




\section{Introduction}

Holography \cite{Maldacena:1997re,Gubser:1998bc,Witten:1998qj} defined
by equality of partition functions $Z_{CFT}=Z_{AdS}$ alongwith the
Ryu-Takayanagi formulation
\cite{Ryu:2006bv,Ryu:2006ef,Hubeny:2007xt,Rangamani:2016dms}
beautifully encodes nice quantum entanglement properties of boundary
CFTs and their subregions via geometric properties of associated bulk
extremal surfaces in $AdS$. We expect that more realistic
gravitational systems like de Sitter space distort these nice
features: with the (spatial) holographic screen at future timelike
infinity $\scri^+$ taken as the natural boundary, $dS/CFT$
\cite{Strominger:2001pn,Witten:2001kn,Maldacena:2002vr,Anninos:2011ui}
suggests more exotic ghost-like non-unitary Euclidean CFT duals\ (we
expect ordinary Euclidean CFTs are dual to Euclidean $AdS$).

Studies of RT/HRT extremal surfaces anchored at $\scri^+$
\cite{Narayan:2015vda,Sato:2015tta,Narayan:2015oka,Narayan:2017xca,
  Narayan:2019pjl,Narayan:2020nsc} and recent reinventions
\cite{Doi:2022iyj,Narayan:2022afv,Doi:2023zaf,Narayan:2023ebn,
  Narayan:2023zen,Goswami:2024vfl,Nanda:2025tid} amount to bulk
analogs of ``boundary entanglement entropy'' in the dual CFT. There
are no real $\scri^+\ra \scri^+$ turning points
\cite{Narayan:2015vda}: the resulting complex or timelike extremal
surfaces that necessarily arise have complex areas, best interpreted
as pseudo-entropies.  The non-unitary structures in such a CFT imply
that adjoints of states $|\psi\ran$ are nontrivial so that
$\rho_A={\rm Tr}_B(\psi\ran\lan\psi|)$ is akin to a reduced transition
matrix (see \eg\ toy models in \cite{Narayan:2016xwq,Jatkar:2017jwz}).
Thus such a boundary entanglement entropy is best classified as a
pseudo-entropy \cite{Nakata:2020luh}, the entropy based on the
transition matrix $|f\ran\lan i|$ regarded as a generalized density
operator. In general pseudo-entropies are complex-valued with no
positivity properties. From the bulk perspective, the $dS/CFT$
dictionary $Z_{CFT}=\Psi_{dS}$ \cite{Maldacena:2002vr} equates the
Hartle-Hawking Wavefunction of the Universe \cite{Hartle:1983ai} at
late times (with appropriate early time regularity) with the partition
function of the hypothetical dual CFT. This implies that a boundary
replica on $Z_{CFT}$ translates to a bulk replica on the Wavefunction
$\Psi_{dS}$ regarded as a transition amplitude for creating this
universe \cite{Narayan:2023zen}. Indeed explicit bulk $dS$-like
replica geometries with replica boundary conditions at $\scri^+$
vindicate this \cite{Nanda:2025tid}: they enable explicit evaluation
of boundary Renyi$_n$ (pseudo-)entropies whose $n\ra 1$ limits then
become the extremal surface areas.

In this note, we revisit no-boundary $dS$ extremal surfaces and their
areas as pseudo-entropies, for generic subregions at $\scri^+$,
filling some gaps in previous work.  For large subregions,
timelike+Euclidean extremal surfaces exist with transparent geometric
interpretations as do complex ones, focussing on $dS_3$ for
simplicity.  The situation for small subregions is analogous to
Poincare $dS$ \cite{Narayan:2015vda} and only complex extremal
surfaces exist (sec.~\ref{sec:dS3smallsub}). The extremal surface area
integrals are defined via time contours in the complex time plane.  We
find multiple extremal surfaces with indistinguishable areas: their
time contours can be deformed into each other in the complex time
plane without obstruction, so they are equivalent for these
pseudo-entropy purposes (sec.~\ref{sec:dS3cmplxTimeC}).  There are
parallels with complex extremal surfaces in the study of timelike
entanglement,
\eg\ \cite{Heller:2024whi,Milekhin:2025ycm,Guo:2025pru,Das:2023yyl,Nath:2024aqh,Katoch:2025bnh,Heller:2025kvp,Zhao:2025zgm,Fujiki:2025rtx,Li:2025tud,Guo:2025ase,Afrasiar:2025eam,Kanda:2026jyk,Hikida:2022ltr,Li:2022tsv,Jiang:2023ffu,Jiang:2023loq,Chu:2023zah,Chen:2023sry,Chen:2023eic,He:2023ubi,Guo:2024lrr,Fareghbal:2024lqa,Xu:2024yvf,Afrasiar:2024ldn,Harper:2025lav}.
This also suggests equivalences between the complex replica geometries
in \cite{Nanda:2025tid}\ (sec.~\ref{sec:dS3cmplxGeom}). Looking at
multiple subregions, it appears that subregion duality (geometrically)
and entropy inequalities in de Sitter encode $AdS$ ones via analytic
continuation (sec.~\ref{sec:dS3multsub}): this is consistent with
previous discussions in \cite{Narayan:2023zen}. Likewise higher
dimensional $dS_{d+1}$ also shows the $dS$ entropy term arising
nontrivially along complex time contours for small subregions
(sec.~\ref{sec:dSd+1}). We also discuss mapping future boundary
subregions and those on constant time slices in the static patch via
lightrays in entirely Lorentzian $dS$ (sec.~\ref{sec:dSstatscri+}).
Then, analogous to future-past extremal surfaces connecting
$\scri^\pm$, we exhibit left-right extremal surfaces connecting small
subregions in the $N/S$ static patches: their area equals $dS$
entropy.  Sec.~\ref{sec:Disc} has a Discussion.

\section{de Sitter extremal surfaces, generic subregions}\label{sec:dSnbRev}

$AdS$ exhibits a natural optimization: RT/HRT extremal surfaces anchored
at the boundary on constant time slices dip into the bulk radially (to
lower area) upto a ``turning point'' where they begin to return to the
boundary \cite{Ryu:2006bv,Ryu:2006ef,Hubeny:2007xt,Rangamani:2016dms}.
In $dS$ the $\scri^+$ boundary is spatial with no time.
Operationally, we pick some spatial direction to define a boundary
Euclidean time slice on which we define boundary subregions, whose
boundaries serve as anchors for bulk extremal surfaces dipping into
the holographic, time, direction. Extremization shows there are no real
turning points where surfaces starting at $\scri^+$ begin to return to
$\scri^+$.  In Poincare $dS$, it turns out that the only extremal
surfaces that anchor at the boundary for generic subregions and return
are complex ones amounting to analytic continuations from $AdS$
Poincare ones \cite{Narayan:2015vda,Sato:2015tta,Narayan:2015oka}:
their complex areas resemble entanglement entropies with the
exotic central charges in $dS/CFT$. For maximal subregions in global
no-boundary $dS$, the bulk extremal surfaces are ``vertical''
timelike surfaces (pure imaginary area) in the Lorentzian region
which join with spatial surfaces going around the Euclidean hemisphere
\cite{Doi:2022iyj,Narayan:2022afv}, with total area
\be\label{nbdS43rev}
(dS_4)\quad 
S_A = i\,{\pi l^2\over 2G_4} {R_c\over l} + {\pi l^2\over 2G_4}\,;
\qquad\quad
(dS_3)\quad 
S_A = i{l\over 2G_3}\log {2R_c\over l} 
+ {\pi l\over 4G_3}\,.
\ee
The real piece is precisely half de Sitter entropy from the 
hemisphere, but does not directly map to the cosmological horizon
area \cite{Gibbons:1977mu}. The imaginary divergent terms contain
$R_c$, the cutoff
near $\scri^+$. For $dS_3$, this pertains to the maximal subregion
(half-circle) on some equatorial $S^1$ slice of the $S^2$ at $\scri^+$\
(for $dS_4$ we have a hemisphere of the $S^2$ slice in $S^3\in \scri^+$).
These areas can be realized via $AdS$ analytic continuations
$L_{AdS}\ra il$ amounting to space-time rotations.
They can also be recovered by a bulk replica on the Wavefunction
$\Psi_{dS}$ (regarded as a transition amplitude for creating this
universe \cite{Narayan:2023zen}) which is a boundary replica on
$Z_{CFT}$.  Explicit bulk $dS$-like replica geometries (analogs of
\cite{Lewkowycz:2013nqa,Hung:2011nu,Casini:2011kv}) with replica
boundary conditions at $\scri^+$ vindicate this \cite{Nanda:2025tid},
enabling evaluation of boundary Renyi$_n$ (pseudo-)entropies (whose
$n\ra 1$ limits give (\ref{nbdS43rev})). See also
\cite{Anastasiou:2025rvz,Anastasiou:2026bbf} for related discussions.

\subsection{$dS_3$ and generic subregions}\label{sec:dS3smallsub}

\begin{figure}[h] 
  \includegraphics[width=24pc]{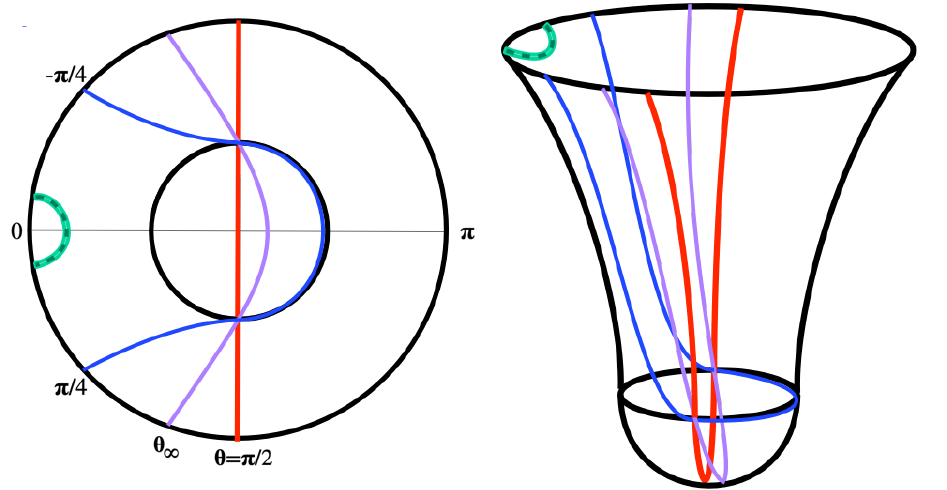} 
\begin{minipage}[b]{15pc}
  \caption{{\label{dS3surftE}\!\! \footnotesize{$dS_3$ no-boundary
        extremal surfaces on an $S^2$ equator slice (right side;\
        left is the ``top view'' from $\scri^+$): the red IR extremal
        surface for maximal $\scri^+$ subregion
        $[-{\pi\over 2}, {\pi\over 2}]$\ \ ($S^{1}$ hemisphere),\ the
        tilted violet curve for non-maximal subregion
        $[-\theta_\infty, \theta_\infty]$,\ 
        the blue limiting timelike+Euclidean surface for 
        $[-{\pi\over 4}, {\pi\over 4}]$,\ and the green dashed 
        complex extremal surface for smaller subregions. }  }}
\end{minipage}
\end{figure}

As the subregion size decreases to say some polar arc, the vertical
surface tilts, as does the hemisphere one, and their joining is more
strained. Eventually for a sufficiently small subregion, the
hemisphere part hits a limiting tilt and then stops existing as a real
spatial surface (see Figure~\ref{dS3surftE}). While these connected
timelike+Euclidean extremal surfaces do not exist, we expect that some
surface must exist whose area gives the boundary entanglement
(pseudo-) entropy for sufficiently small subregions. Since a small
portion of $\scri^+$ resembles the Poincare slicing of $dS$, it would
appear that these are complex extremal surfaces as in those cases
\cite{Narayan:2015vda}. Studying this explicitly vindicates this,
with $dS_3$ being particularly simple. Using the expressions in
\cite{Narayan:2023zen}, the area functional on an equatorial $S^1\in
S^2$ at $\scri^+$ is
\be\label{areafnaldS3}
S = {2\over 4G_3} \int_{r_*}^{R_c} dr\,\sqrt{ -{1\over  {r^2\over l^2}-1}
  + r^2 (\theta')^2} \ \ \xleftrightarrow{\ il\ra L\ }\ \
{2\over 4G_3} \int_{r_*}^{R_c}
dr\,\sqrt{{1\over 1+{r^2\over L^2}} + r^2 (\theta')^2}\ .
\ee
The right side expression is in $AdS_3$ via analytic
continuation. The maximal subregion is the half-circle
$\theta=[-{\pi\over 2},{\pi\over 2}]$: with $\theta=const$, the
turning point is the no-boundary point (nbp) $r=0$ and the area
gives (\ref{nbdS43rev}). For finite subregions, extremization 
gives ${r^2\theta'\over\sqrt{...}}=A$\ so
\bea\label{theta(r)dS3}
(\theta')^2 = {1/r^2\over (1+{r^2\over L^2})({r^2\over A^2}-1)}\,,
&& \  \tan\big(\theta-{\pi\over 2}\big) =  
           {\sqrt{1+{r^2\over L^2}}\over \sqrt{{r^2\over A^2}-1}}\,,
\qquad\qquad\quad [AdS] \nn\\
(\theta')^2 = {1/r^2\over (1-{r^2\over l^2})({r^2\over A^2}-1)}\,,
&& \ \tan\big(\theta-{\pi\over 2}\big)
= {\sqrt{1-{r^2\over l^2}}\over\sqrt{{r^2\over A^2}-1}}\,,
\qquad\quad  [dS: r<l] \nn\\  
(\theta')^2 = {1/r^2\over ({r^2\over l^2}-1)(1+{r^2\over A^2})}\,,
&& \ \tan\big(\theta-{\pi\over 2}\big)
= -{\sqrt{{r^2\over l^2}-1}\over \sqrt{{r^2\over A^2}+1}}\,,
\qquad\ [dS: r>l].\quad    
\eea
These are related by the analytic continuation\ $L^2\ra -l^2,\ A^2\ra A^2$\
for $AdS$ to $dS,\ r<l$ (hemisphere) and\ $L^2\ra -l^2,\ A^2\ra -A^2$\
for $AdS$ to $dS,\ r>l$ (Lorentzian). The asymptotics are
\be\label{dS3asympAtheta_infty}
\theta\xrightarrow{r\ra\infty} {\pi\over 2} - \tan^{-1}{A\over l}
\equiv \theta_\infty\,;\qquad \theta\xrightarrow{r\ra l} {\pi\over 2}\,;
\qquad \theta\xrightarrow{r\ra A} \pi\,.
\ee
This describes the $dS$ surface anchored at $\theta_\infty\in I^+$
going into the time direction as a timelike surface, hitting the $r=l$
slice at $\theta={\pi\over 2}$ and then going around the hemisphere
till $\theta=\pi$ at the turning point $r=A$\ (note that the
hemisphere in Figure~\ref{dS3surftE} should be understood as living in
the Euclidean direction). This then joins with a similar half-surface
on the other side (of $\theta=0$, so $\theta<0$ now) going from
$\theta=\pi$ at $r=A$ to $\theta=-{\pi\over 2}$ at $r=l$ and thence
to $-\theta_\infty\in I^+$. The boundary subregion at $\scri^+$ is
$[-\theta_\infty,0]\cup[0,\theta_\infty]$, with width
$\Delta\theta=2\theta_\infty$ defined by $A^2>0$ as above. It is worth
noting that the joining of the timelike Lorentzian surface to the
Euclidean one (even without tilt) at the complexification point $r=l$
is nontrivial and could have ambiguities: in the above our joining
prescription (smooth $(\theta')^2$) is equivalent to requiring that
the joining is smooth and umabiguous under the $AdS$ analytic
continuation (\ref{theta(r)dS3})\ (amounting to a space-time
rotation).\ The total area of these surfaces becomes 
\be\label{dS3areaTimelike+Eucl}
S_t=S_t^{r>l}+S_t^{r<l} = i{l\over 2G_3}\log{2R_c\over l}
+ i{l\over 4G_3}\log(\sin^2\theta_\infty) + {\pi l\over 4G_3}\,,
\ee
The hemisphere $r<l$ surface in (\ref{theta(r)dS3}) shows that
\be\label{dS3tElimit}
(\theta')^2>0\quad \Rightarrow\quad 0\leq A\leq r\leq l
\quad\Rightarrow\quad
{\pi\over 2}\geq \theta_\infty\geq {\pi\over 4}\,,
\ee
with the $\theta_\infty$ values from (\ref{dS3asympAtheta_infty}).
The turning point is $r_*=A=0$ for the IR surface. As $A$
increases, the hemisphere surface (roughly) tilts upward. The
condition $A\leq l$ is nontrivial: as $A\ra l$, \ie\
$2\theta_\infty\ra {\pi\over 2}$, we obtain from (\ref{theta(r)dS3})
\be
A\ra l:\quad
\tan(\theta-{\pi\over 2})\ra \sqrt{1-r^2/l^2\over r^2/l^2-1}=\pm i\,.
\ee
This indicates a breakdown of the hemisphere surface beyond this point
($\Delta\theta\equiv 2\theta_\infty<{\pi\over 2}$ from
(\ref{dS3asympAtheta_infty}), \ie\ angular subregions spanning less
than a quadrant of the $S^1$ \cite{Narayan:2023zen}), and thus of
these timelike+Euclidean connected (real) extremal surfaces. This
is depicted in Figure~\ref{dS3surftE}.

\noindent However even for smaller subregions, one would expect that some
surfaces anchored at $\scri^+$ must exist whose area encodes boundary
entanglement entropy for these sub-maximal $\scri^+$ subregions.
Focussing on the Lorentzian region $r>l$, extremizing (\ref{areafnaldS3})
gives
\be\label{dS3theta(r)cmplx}
(\theta')^2 = {1/r^2\over ({r^2\over l^2}-1)(1-{r^2\over A^2})}\ \
\xrightarrow{\ r= i\rho\,,\ \ A=i\al }\ \
\Big({d\theta\over d\rho}\Big)^2
= {1/\rho^2\over (1+{\rho^2\over l^2})({\rho^2\over \al^2}-1)}\ .
\ee
The expression on the left can be understood as the $(\theta')^2$
extremization equation in the second line in (\ref{theta(r)dS3}),
rewritten to make $r>l$ manifest, with $A^2$ unchanged. (In contrast
with the third line in (\ref{theta(r)dS3}), this approximates for
large $r$ to\ $(\theta')^2 = -{l^2 A^2\over r^6}$\ which has no
real-valued interpretation: more on this below.)
The expression on the right in (\ref{dS3theta(r)cmplx}) above is
obtained along the complex time path $r=i\rho$ with the turning point
also complexified, to $\rho_*=\al$.
This is now identical to the $AdS$ extremization, the complexification
effectively taking us to an auxiliary $AdS$ space. The solution is
\be\label{dS3CmplxAsympAtheta_infty}
\tan\big(\theta-{\pi\over 2}\big)
= -{\sqrt{1+{\rho^2\over l^2}}\over \sqrt{{\rho^2\over\al^2}-1}}\ ;
\qquad
\theta\xrightarrow{\rho\ra\infty} {\pi\over 2} - \tan^{-1}{\al\over l}
\equiv \theta_\infty\,,\quad
\theta \xrightarrow{\,\rho\ra \al\,} 0\,.
\ee
Smaller
width $\Delta\theta=2\theta_\infty$ give larger $\al=l/\tan\theta_\infty$
\ie\ turning point closer to the auxiliary $AdS$ boundary. 
Note that near the turning point $\al$, these surfaces approach
$\theta=0$ in contrast with (\ref{dS3asympAtheta_infty}) for the
timelike+Euclidean surfaces\ (Figure~\ref{dS3surftE}). 
Note also that although the time contour is complex (imaginary time
path $r=i\rho$), the functional expression for $\theta(r)$ is
real-valued: so the dual CFT spatial subregion size is interpreted
as real. The area for these complex extremal surfaces is
\bea\label{dS3areaCmplx}
S = i\,{2\over 4G_3} \int_\al^{\rho_c} d\rho\sqrt{{1\over 1+{\rho^2\over l^2}}
  \Big({1\over 1-{\al^2\over\rho^2}}\Big)}
&=& -i{l\over 2G_3}
\log(\sqrt{2(\rho^2+l^2)}-\sqrt{2(\rho^2-\al^2)})\Big\vert_\al^{\rho_c}
\nn\\
&=& i{l\over 2G_3}\log{2R_c\over l} + i{l\over 4G_3}\log(\sin^2\theta_\infty)
+ {\pi l\over 4G_3}\,,\qquad
\eea
expanding out and using (\ref{dS3CmplxAsympAtheta_infty}) to obtain
the finite part. Note that the surface lives entirely in the
complexified Lorentzian part of $dS$. There is no manifest hemisphere
interpretation here: the $dS$ entropy term remarkably arises as\
$i{l\over 2G_3}\,\log(-i)$ using $\rho_c=-iR_c$.

These complex extremal surfaces defined along complex time paths
$r\ra ir$ are analogs of those in \cite{Narayan:2015vda} for Poincare
$dS_{d+1}$ with metric $ds^2={R_{dS}^2\over\tau^2}(-d\tau^2+dx_i^2)$: we
now review this. A strip subregion on some boundary Euclidean time
$w=const$ slice of $I^+$ with width along $x$ \ (with $w$, $x$ any of
the $x_i$) gives the area functional and thereby a bulk extremal
surface $x(\tau)$ described by
\be\label{PoincaredS}
S={R_{dS}^{d-1}V_{d-2}\over 4G_{d+1}}\int
{d\tau\over\tau^{d-1}}\sqrt{{\dot x}^2-1}\,,\qquad
{\dot x}^2 = {-B^2\tau^{2d-2}\over 1-B^2\tau^{2d-2}} \qquad
\Big[{\dot x}\equiv {dx\over d\tau}\Big] .
\ee
A turning point where the surface starting at $I^+$ begins to turn
back requires $|{\dot x}|\ra\infty$ while here $|{\dot x}|\leq 1$ for
real $\tau$ and $B^2<0$. For $B^2>0$ the denominator appears to
indicate unbounded growth for $|{\dot x}|$ but note that\
${\dot x}^2 \sim -B^2\tau^{2d-2}$\ near $\scri^+$ where $\tau\sim 0$,
so these are not real surfaces for real $\tau$.\
Complex extremal surfaces with real width $\Delta x$ arise along
complex time paths $\tau\ra i\tau$ and amount to analytic
continuations from $AdS$, leading to areas that resemble boundary
entanglement entropies in $dS/CFT$
\cite{Narayan:2015vda,Sato:2015tta,Narayan:2015oka}. For instance
$dS_3$ above gives\
${\dot x}^2={-B^2\tau^2\over 1-B^2\tau^2}$\ so\
$\tau\ra i{\tilde\tau},\ B^2\ra -{\tilde B}^2$,\ gives a turning point
at $\tau_*\ra i{\tilde\tau}_*={i\over {\tilde B}}$\,. This gives area\
$S\sim i{R_{dS}\over G_3}\log {l\over\epsilon}$ which resembles the
familiar Calabrese-Cardy expression\ ${c\over 3}\log {l\over\epsilon}$\
with the imaginary central charge of $dS_3/CFT_2$ \cite{Maldacena:2002vr}.

An alternative way to see this is to note that the vicinity of the
future boundary in global $dS$ locally resembles that in Poincare
$dS$ with planar foliations since the sphere curvature is negligible.
Explicitly, the metric for large $r$ approximates as
\be\label{dSglobal-Poincare}
ds^2=-{dr^2\over {r^2\over l^2}-1}+r^2d\Omega_d^2\quad
\xrightarrow{\ r\gg l\ }\quad
{l^2\over\tau^2}(-d\tau^2+l^2\Omega_d^2)\,,\qquad
\Big[\tau={l^2\over r}\Big]\,,
\ee
which is locally planar.
The extremization (\ref{dS3theta(r)cmplx}) becomes\
$(\theta')^2 \sim {1/r^2\over ({r^2/l^2})(1-{r^2/A^2})}$\
for $r\gg l$, which can be mapped via $r\ra ir,\ A^2\ra -A^2$,
to the above Poincare case.


\subsection{Complex time plane: time contours and deformations}
\label{sec:dS3cmplxTimeC}

We have seen that for small subregion size the timelike+Euclidean
surfaces do not exist (Figure~\ref{dS3surftE}) but complex extremal
surfaces (\ref{dS3theta(r)cmplx}) do: their area (\ref{dS3areaCmplx})
is identical structurally to (\ref{dS3areaTimelike+Eucl}), but notably
the real $dS$ entropy term arises via complexification with no direct
hemisphere interpretation.

\begin{figure}[h] 
\hspace{1.5pc}
  \includegraphics[width=12pc]{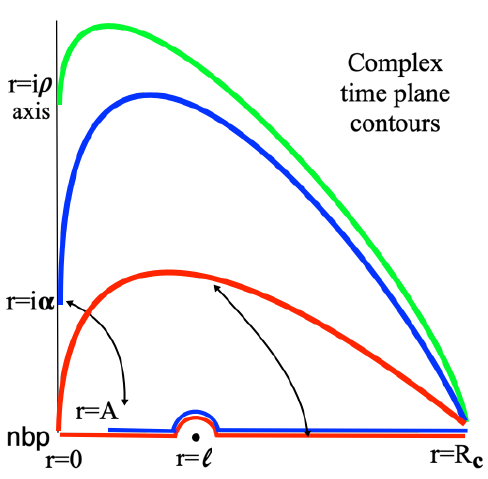} 
\hspace{3pc}
\begin{minipage}[b]{19.2pc}
  \caption{{\label{TimeContours}\!\! \footnotesize{Time contours in
        the complex time $r$-plane for timelike+Euclidean (red or
        blue) and complex $dS_3$ extremal surfaces (IR = red, large
        subregion = blue,\ small subregion = green, with turning
        points on the imaginary-$r$ axis), and deformations thereof
        (skirting the potential pole at $r=l$).\\ } }}
\end{minipage}
\end{figure}
It appears that these complex surfaces exist for all boundary
subregions, including large ones where $\theta_\infty>{\pi\over 4}$\,.
So an obvious question when both exist, is: are these extremal
surfaces distinct, and if so how do we choose which one is the
correct one? To make this sharp, consider maximal subregions and
the IR extremal surface in the two cases:\\
(1) timelike+Euclidean extremal surfaces (\ref{theta(r)dS3}) that
exist as real curves can be drawn on the spacetime/Penrose diagram
of $dS_3$ (colloquially speaking). They can be obtained via analytic
continuation $L_{AdS}\ra il$ amounting to a space-time rotation
from $AdS$ \cite{Narayan:2023zen}. Their area is
\be\label{dS3IR-tE}
S_{tE} 
= i\,{2\over 4G_3} \int_l^{R_c} {dr\over\sqrt{{r^2\over l^2}-1}} \, +\,
{2\over 4G_3} \int_0^{l} {dr\over\sqrt{1-{r^2\over l^2}}}
= i\,{l\over 2G_3} \log{2R_c\over l} + {\pi l\over 4G_3}
\,,
\ee
using (\ref{areafnaldS3}) with $\theta'=0$ (or
(\ref{dS3areaTimelike+Eucl}) with $\theta_\infty={\pi\over 2}$). This
is the time contour $r=[0,l]\cup [l,R_c]$ in the complex-$r$ time
plane, with $r=0$ the no-boundary point (nbp), $r=l$ the complexification
point and $r=R_c$ the cutoff at $\scri^+$: see Figure~\ref{TimeContours}.
There is a
potential pole at $r=l$ where the integrand can develop a singularity
due to the vanishing denominator: in practice however the integral
itself is nonsingular for pure $dS_{d+1}$.\\
(2) Complex extremal surfaces (\ref{dS3theta(r)cmplx}),
(\ref{dS3CmplxAsympAtheta_infty}), live in the auxiliary $AdS$ space
complexifying from $dS$: they cannot be drawn as curves in the $dS$
spacetime/Penrose diagram. Their area is
\be\label{dS3IR-cmplx}
S_{i\rho} = i\,{2\over 4G_3} \int_0^{\rho_c}
{d\rho\over\sqrt{1+{\rho^2\over l^2}}}
= i{l\over 2G_3}\log{2\rho_c\over l} 
= i{l\over 2G_3}\log{2R_c\over l} + {\pi l\over 4G_3}\,,
\ee
from (\ref{dS3areaCmplx}) for $\al=0$, with $\rho_c=-iR_c$ giving
$i{l\over 2G_3}\,\log(-i)=i{l\over 2G_3}({-i\pi\over 2})$ as the
real term (choosing this branch of the logarithm).
This is a $-EAdS$ time contour $r=[0,R_c]$ in the complex-$r$ time
plane. For any nonmaximal subregion the turning point is at $r=i\al$
on the imaginary $r$-axis, so the maximal subregion turning point
$r=0$ is approached along imaginary time paths (so we will call this
the imaginary time contour). Strictly speaking however, the area
(\ref{dS3IR-cmplx}) as such depends only on the endpoints so the
precise path in the complex time plane does not seem to matter: we
depict this by the general time contour curves in
Figure~\ref{TimeContours}\ (similar features appear in higher dim
$dS$, sec.~\ref{sec:dSd+1}).
For arbitrary complex-$\rho$, the surfaces (\ref{dS3theta(r)cmplx}),
(\ref{dS3CmplxAsympAtheta_infty}), are complex functions $\theta(r)$
anchored at real $\theta$-subregions of $\scri^+$: for $r=i\rho$
(\ie\ real $\rho$), $\theta(\rho)$ is a real-valued function.
If we consider perturbations of de Sitter space, it is likely that
general complex extremal surfaces in the complex-$\rho,\theta$ space
must be allowed and will play crucial roles (see
\cite{Fujiki:2025rtx}): we will not discuss this further here.

In general in $AdS$, when multiple RT/HRT saddles exist, we pick the
lower area one. In the present $dS_3$ case, the areas (\ref{dS3IR-tE})
and (\ref{dS3IR-cmplx}) are identical. These IR surfaces for maximal
subregions are perhaps the simplest examples of this kind: generic
large subregions also exhibit similar behaviour as we have seen (see also
\cite{Heller:2024whi,Milekhin:2025ycm,Guo:2025pru,Das:2023yyl,Nath:2024aqh,Katoch:2025bnh,Heller:2025kvp,Zhao:2025zgm,Fujiki:2025rtx,Li:2025tud,Guo:2025ase,Afrasiar:2025eam,Kanda:2026jyk,Hikida:2022ltr,Li:2022tsv,Jiang:2023ffu,Jiang:2023loq,Chu:2023zah,Chen:2023sry,Chen:2023eic,He:2023ubi,Guo:2024lrr,Fareghbal:2024lqa,Xu:2024yvf,Afrasiar:2024ldn,Harper:2025lav}
in the study of timelike entanglement). So we now ask if these
surfaces must somehow be considered equivalent.  This is reasonable if
the time contours (1) and (2) can be deformed into each other in the
complex-$r$ time plane: and indeed this is so.  The only potential
pole is at $r=l$ so there is in fact no obstruction to deforming the
imaginary time $r=i\rho$ contour (2) into the timelike+Euclidean
$r=[0,l]\cup [l,R_c]$ time contour in the upper half plane above
$r=l$.
This contour deformation argument appears reasonable when multiple
equivalent surfaces/contours exist for the same boundary subregion (of
course for small subregions only the complex ones exist). However it
is worth noting that the timelike+Euclidean time contour admits a
transparent geometric interpretation directly in $dS_3$, while the
complex time contour, living in an auxiliary $AdS$ space, 
acquires geometric interpretation after contour deformation.

The above description of identifying extremal surfaces whose time
contours are deformable in the complex time plane is consistent with
the discussions of no-boundary extremal surfaces in situations like
slow-roll inflation regarded as deformations away from $dS$.  We refer
to \cite{Goswami:2024vfl} for detailed discussions of no-boundary
slow-roll IR extremal surface areas for maximal subregions and the
Wavefunction of the Universe to $O(\epsilon)$.  In these cases, the
extremal surface area integrals must be defined in terms of
appropriate time contours in the complex time plane going around the
pole at the complexification point $r=l$ where singularities do occur:
this then enables evaluation of the $O(\epsilon)$ corrections to the
$dS$ areas.  The details of the contour regularization are unimportant
as long as the pole is avoided. However it is worth noting that a
large departure from the timelike+Euclidean contour appears to not
make sense there. For instance, the $-EAdS$ contour does not appear to
yield the same extremal surface area with corresponding $O(\epsilon)$
corrections to the entropy. This is perhaps not surprising since it
turns out that the inflaton+metric perturbations in no-boundary
slow-roll inflation \cite{Goswami:2024vfl} do not admit any simple
$AdS$ interpretation under analytic continuation, nor do the IR
extremal surface areas correspondingly, which appear best understood
via the manifestly geometric timelike+Euclidean contours. This
suggests that the $-EAdS$ time contour is likely to make sense only
when there exists a reasonable auxiliary $AdS$ interpretation for the
geometries and the corresponding extremal surfaces. Thus overall, for
perturbations away from pure de Sitter and cosmological situations
more generally, extremal surface time contours will likely exhibit
nontrivial features.\\
For $dS_3$ slow-roll inflation, near the future boundary we
have\ $ds^2 = -{1\over r^2}(1+\epsilon\log r) dr^2 + r^2d\theta^2$\ on
an equatorial $S^2$ slice at $\scri^+$. Considering small subregions
now, in light of our discussion for $dS_3$, we expect that only
complex extremal surfaces exist here (\ie\ no timelike+Euclidean
ones). Setting up the area functional for a small $\theta$-arc
subregion at $\scri^+$ appears to confirm this via Mathematica but
closed form expressions for the $O(\epsilon)$ corrections seem
difficult.

\subsection{Complex replica geometries and equivalences}\label{sec:dS3cmplxGeom}

The a priori distinct area integrals and time contours that are
equivalent under deformations in the complex time plane imply
equivalences between a priori distinct complex replica geometries
correspondingly. The area (\ref{dS3IR-tE}) corresponds to the replica
geometry \cite{Nanda:2025tid}
\be\label{dS3replicaSdS3}
ds^2 = - {d\varrho^2\over {\varrho^2\over l^2} - {1\over n^2}} +
\Big({\varrho^2\over l^2} - {1\over n^2}\Big) dt^2 + \varrho^2 d\varphi^2\,,
\qquad \varphi\equiv \varphi+2\pi n\,,
\ee
in the Lorentzian region. The Euclidean part is obtained from the
$\varrho<{l\over n}$ part above with Euclidean time $t\ra i\tau_E$ and
$\tau_E=[0,{\pi\over 2}]$.\
The replica boundary conditions are imposed on the holographic screen
at large real $\varrho=R_c\gg l$.\ 

By comparison, the area (\ref{dS3IR-cmplx}) is from a $-EAdS$ contour.
In this regard $dS_{d+1}$ replica geometries with hyperbolic foliations
(with $dH_{d-1}^2=d\chi^2+\sinh^2\chi d\Omega_{d-2}^2$) were studied for
$d\geq 4$ in \cite{Nanda:2025tid}: these are closely related to
\cite{Hung:2011nu,Casini:2011kv} in $AdS$. For the $dS_3$ case this
gives
\be\label{dS3replicaCmplx}
ds^2 = - {dr^2\over {r^2\over l^2} + {1\over n^2}} +
l^2\Big({r^2\over l^2} + {1\over n^2}\Big) d\phi^2 + r^2 d\chi^2\,,
\qquad \phi\equiv \phi+2\pi n\,.
\ee
The IR extremal surface area at $n=1$ maps to the area of the
complex horizon at $r_h=il$,
\be
S_{IR} = {r_h\,(2V_\chi)\over 4G_3}  =
{i l\over 2G_3} \int_0^{\chi_{max}} d\chi = 
i{l\over 2G_3} \log{2R_c\over l} + {\pi l\over 4G_3}\,,\qquad
\chi_{max}\sim \log {2R_c\over il}\,.
\ee
The nontrivial $\chi_{max}$ value arises since this geometry (for
$n=1$) is a nontrivial embedding into global $dS_3$ given by\
$ds^2=-{d\hat{r}^2\over \hat{r}^2-1}+\hat{r}^2(d\al^2+\sin^2\al\,d\beta^2)$.
Using embedding coordinates in global and hyperbolic foliations,
adapting from \cite{Nanda:2025tid}, gives
\be\label{chimax}
{1\over\cosh\chi} = \sqrt{\frac{\hat{r}^2 \sin^2\alpha-l^2}{\hat{r}^2-l^2}}\ ,
\quad r=\sqrt{\hat{r}^2 \sin^2\alpha-l^2}\quad\ra\quad
{1\over\cosh\chi_{max}} \sim \frac{i l}{R_c}\,.
\ee
The global subregion range $\alpha \in [0,\frac{\pi}{2}]$ translates
to complex $\chi$ values in the Lorentzian part $\hat{r}>l$: for
$\hat{r}=R_c$ with $\sin\al < {l\over R_c}$ we obtain, as $\al\ra 0$,
the limiting (imaginary) value $\chi_{max}$ above. On the other
hand, as $\al\ra {\pi\over 2}$ we see that $\chi\ra 0$.
This results in the above complex area of the complex horizon.

The fact that the IR extremal surface area for maximal subregions
arises from the timelike+Euclidean curve in (\ref{dS3replicaSdS3}) and
from the complex horizon in (\ref{dS3replicaCmplx}) suggests that
these geometries are equivalent in some sense, for these
pseudo-entropy purposes here. (Note that the corresponding $AdS$
replica geometries correspond simply to different $AdS$ foliations,
with the extremal surfaces arising via either cosmic branes as in
\cite{Lewkowycz:2013nqa} or hyperbolic $AdS$ black hole horizons as
in \cite{Hung:2011nu}.) The nontrivial embedding coordinate relations
into global $dS_3 \equiv (\hat{r},\al,\beta)$ for $n=1$ for both
geometries suggests complex coordinate transformations
$\{(\varrho,t,\varphi) \leftrightarrow (r,\phi,\chi)\}$ via (\ref{chimax})
of the geometries as a whole, not just in the complex-$r$ time plane.
It would be interesting however to understand better how these are
systematized. More complicated subregions and higher dimensional cases
are likely to have more intricate features, but these are harder to
find explicitly beyond the ones in \cite{Nanda:2025tid}.

Finally it is worth noting that the $dS$ replicas here are complex
geometries. We demand that the metric Euclideanized as $r\ra i\rho$
satisfy regularity in the interior \cite{Nanda:2025tid}, along the
lines in \cite{Das:2013mfa} (see also \cite{Maldacena:2019cbz}),
similar to regularity criteria employed in asymptotically $AdS$
spaces. This is a priori different from the KSW criterion
\cite{Kontsevich:2021dmb,Witten:2021nzp}, although similar in spirit.

\subsection{ Multiple small subregions}\label{sec:dS3multsub}

The complex $dS_3$ surfaces above involve analytically continuing
$r\ra ir$ from $AdS$. While they associate ``bounded'' bulk subregions
to small boundary subregions, they live in the auxiliary $AdS$ space.
Subregion duality, just geometrically, thus operates there. Entropy
inequalities here associated to multiple subregions are also
complex-valued: they encode the familiar $AdS/CFT$ entropy
inequalities via analytic continuation.

This was observed for large subregions in \cite{Narayan:2023zen} (in
particular pseudo-entropy analogs of mutual information, tripartite
information and strong subadditivity) via the timelike+Euclidean
surface (complex) areas but similar features hold for small subregions
as well. To see this, consider $dS_3$ and two small disjoint
$\theta$-arc subregions on the equatorial $S^1$, defined by
$[\theta_1,\theta_2]$ and $[\theta_3,\theta_4]$, where each width
is small. Since only complex extremal surfaces in the auxiliary
$AdS$ space exist for these, we expect their properties resemble
those in $AdS$. In terms of the explicit complex extremal surface
parametrizations (\ref{dS3theta(r)cmplx}),
(\ref{dS3CmplxAsympAtheta_infty}), the
widths $2\theta^{(1)}_\infty=\theta_2-\theta_1$ and
$2\theta^{(2)}_\infty=\theta_3-\theta_2$ and
$2\theta^{(3)}_\infty=\theta_4-\theta_3$ are all small. Then the
disconnected extremal surfaces connect $[\theta_1,\theta_2]$ and
$[\theta_3,\theta_4]$, while the connected extremal surface connects
$[\theta_1,\theta_4]$ and $[\theta_2,\theta_3]$.
The divergent and $dS$ entropy pieces in the areas
(\ref{dS3areaCmplx}) cancel so the difference between the
disconnected and connected surface areas becomes
\be\label{dS3conndisc}
S^{conn}-S^{disc} \sim  
i{l\over 4G_3}\log\Big(
{\theta_{14}^2\,\theta_{23}^2\over \theta_{12}^2\,\theta_{34}^2} \Big)
= i{l\over 2G_3} \log \Big({1-x\over x}\Big)\,,\qquad
x={\theta_{12}\theta_{34}\over\theta_{13}\theta_{24}}\,,
\quad 0<x<1\,,
\ee
where $\theta_{ij}=\theta_i-\theta_j$ etc, and approximating
$\sin^2\theta_{ij}\sim \theta_{ij}^2$ since all $\theta_{ij}$'s are
small.  The above expression resembles the mutual information
expression involving the cross-ratio $x$\ for two intervals
\cite{Headrick:2010zt},\cite{Hayden:2011ag}, after analytic
continuation $il\ra L$ back to $AdS$. Thus we see that the familiar
disentangling transition from connected to disconnected surfaces
in $AdS$ for well-separated intervals ($x<{1\over 2}$) is encoded
through analytic continuation here via the complex surfaces in $dS$\
(see \cite{Harper:2025lav} for related discussions in timelike
entanglement).\ Likewise for three subregions\ $A=[\theta_1,\theta_2],\
B=[\theta_2,\theta_3],\ C=[\theta_3,\theta_4]$,\ using
(\ref{dS3areaCmplx}), the tripartite information simplifies to
(the divergence and the $dS$ entropy terms cancelling)
\be
I_3=S_A+S_B+S_C-S_{AB}-S_{BC}-S_{AC}+S_{ABC}=
i{l\over 2G_3}\log x\,.
\ee
So under $il\ra L$, we have\ $I_3\leq 0$ as is well-known in $AdS$
\cite{Hayden:2011ag}.

For large subregions also, it appears that geometric subregion duality
(boundary subregion $\ra$ bulk subregion defined by boundary and
extremal surface) is not valid directly in the $dS$ geometry, but is
only encoded via the $AdS$ analytic continuations
\cite{Narayan:2023zen} (sec.2.4, 2.5). For instance, the bulk $dS$
subregions overlap even if the boundary subregions are disjoint,
except when the subregions are maximal (see Fig.3, Fig.4 there). This
appears consistent with cases where some subregions are small while
others are large. As a simple example, for\
$A=[-{\pi\over 3},{\pi\over 3}],\ \ B=[-{2\pi\over 3},{2\pi\over 3}]$\ as
``oppositely'' located subregions in the equatorial $S^1$, unconnected
timelike+Euclidean surfaces exist for $A$ and $B$ separately, but
apparently no connected ones which connect $[-{\pi\over 3},-{2\pi\over
    3}]$ and $[{\pi\over 3},{2\pi\over 3}]$\ (using
(\ref{dS3tElimit}), since $\Delta\theta={\pi\over 3}<{\pi\over
  2}$). So the connected surfaces must be complex ones as above: after
various cancellations, the area difference is just in the finite
imaginary parts structurally similar to (\ref{dS3conndisc}) above.

Thus the bulk subregion dual to the boundary subregion in $dS$ is
again defined via analytic continuation from the corresponding ones in
$AdS$. Subregion duality here is not directly manifest in $dS$ but
encodes that in the auxiliary $AdS$ space (this is just geometric,
weaker than entanglement wedge reconstruction).
It would be fascinating however to understand this as well as multiple
subregions and disentangling transitions intrinsically in the
nonunitary CFTs pertinent to $dS/CFT$, possibly via relative
pseudo-entropy and so on.

\subsection{Higher dimensional $dS_{d+1}$}\label{sec:dSd+1}

For higher dimensional global de Sitter $dS_{d+1}$, let us first
examine IR surfaces for maximal subregions. With the timelike+Euclidean
contour, these have area
\cite{Doi:2022iyj,Narayan:2022afv,Narayan:2023zen,Nanda:2025tid}
\bea\label{dSd+1-StE}
S_{tE} &=& S^{r<l} + S^{r>l} = {V_{S^{d-2}}\over 4G_{d+1}} \int_0^l
{r^{d-2}\, dr\over \sqrt{1-{r^2\over l^2}}}
+ i {V_{S^{d-2}}\over 4G_{d+1}} \int_l^{R_c}
{r^{d-2}\, dr\over \sqrt{{r^2\over l^2}-1}} \nn\\
&=& {1\over 2} {l^{d-1}V_{S^{d-1}}\over 4G_{d+1}} +
i\# {l^{d-1}\over 4G_{d+1}} {R_c^{d-2}\over l^{d-2}} +\, \ldots ,
\eea
giving, \eg\ for $dS_4$ and $dS_3$, the areas in (\ref{nbdS43rev}).\
By comparison the complex $-EAdS$ contour with $r=i\rho$ gives the area
\be\label{dSd+1-Sirho}
S_{i\rho} = i^{d-1} {V_{S^{d-2}}\over 4G_{d+1}} \int_0^{\rho_c} 
{\rho^{d-2}\, d\rho\over \sqrt{1+{\rho^2\over l^2}}}
=  i\, i^{d-2} (-i)^{d-2}  
\# {l^{d-1} \over 4G_{d+1}} {R_c^{d-2}\over l^{d-2}} + \ldots
\ee
where $\rho_c=-iR_c$, and in the leading divergence term various $i$-
and $(-i)$-factors cancel, in the end leaving a single overall
$i$. The $dS$ entropy term needs to be obtained case by case. Overall
this area can then be seen to evaluate to the corresponding
timelike+Euclidean surface area. For instance for $dS_4$ and $dS_5$,
the complex $-EAdS$ contour gives
\bea
&& S_{i\rho}^{dS_4} =
-{2\pi l^2\over 4G_4}\sqrt{1+{\rho^2\over l^2}}\Big\vert_0^{\rho_c}
= i{\pi l^2\over 2G_4} {R_c\over l} + {\pi l^2\over 2G_4}\,,\nn\\
&& S_{i\rho}^{dS_5} =
-i{\pi l^3\over 2G_5} \left({\rho_c\over l}\sqrt{1+{\rho_c^2\over l^2}} +
\log{l\over 2\rho_c}\right) = i {\pi l^3\over 2G_5} {R_c^2\over l^2}
+ i {\pi l^3\over 2G_5} \log {2R_c\over l} +  {\pi^2 l^3\over 4G_5}\,,\qquad
\eea
which can be seen to match $S_{tE}$ in (\ref{dSd+1-StE}) above.
The $dS$ entropy terms have no manifest hemisphere interpretation.
For $dS_5$ this arises as\ $i{\pi l^3\over 2G_5}\log(-i)$,\
similar to (\ref{dS3IR-cmplx}) for $dS_3$.

Let us now consider higher dimensional $dS_{d+1}$ and extremal
surfaces for small subregions of the future boundary. Then we can
approximate the metric as (\ref{dSglobal-Poincare}) and consider an
$S^d$ equatorial plane slice and a small polar cap subregion with
latitude angle width $\Delta\theta$.
Extremal surfaces anchor at the $\theta$-latitude boundary of this
polar cap subregion and wrap the $S^{d-2}$:
\be
S = {l^{2d-3}V_{S^{d-2}}\over 4G_{d+1}} \int {d\tau\over\tau^{d-1}}\,
(\sin\theta)^{d-2}\,\sqrt{l^2(\theta')^2-1}\ \sim\
{l^{2d-3}V_{S^{d-2}}\over 4G_{d+1}} \int {d\tau\over\tau^{d-1}}\,
\theta^{d-2}\,\sqrt{l^2(\theta')^2-1}\,,
\ee
is the area functional where, in accord with small polar cap
subregions, we have approximated $\theta$ to be small to obtain the
expression on the right. This we recognize is identical to the
analysis of extremal surfaces for spherical subregions in Poincare $dS$
\cite{Narayan:2015oka}, which we revisit. The extremization equation
and solution (anchored at $l\theta=a$ on $\scri^+$ where $\tau=0$) are
\be
   {d\over d\tau} \left({\theta^{d-2}\over\tau^{d-1}}\,
   {l^2\theta'\over\sqrt{l^2(\theta')^2-1}}\right)
   = (d-2) {\theta^{d-3}\over\tau^{d-1}}\,\sqrt{l^2(\theta')^2-1}
   \quad\ra\quad  l\theta(\tau) = \sqrt{a^2+\tau^2}\,,
\ee
These are in fact analytic continuations
of the familiar extremal surfaces $l\theta(T) = \sqrt{a^2-T^2}$
for spherical subregions in $AdS$. The turning point is in the
complex $\tau$-plane at $\tau=-ia$.\
The area functional on-shell becomes
\be\label{dSd+1areaCmplx}
S = {l^{d-1}V_{S^{d-2}}\over 4G_{d+1}} \int_{C_\tau}    
{i a\,d\tau\over\tau^{d-1}}\ (\sqrt{a^2+\tau^2})^{d-3}\,,
\ee
with $C_\tau$ a time path in the complex $\tau$-plane with
endpoints $\tau_{UV}={l^2\over R_c}\sim 0$ and $\tau_{IR}=-ia$: 
the precise details of the path do not enter in the area, as in
Figure~\ref{TimeContours}.
For $dS_4$ (\ie\ $d=3$) and $dS_5$, this area (\ref{dSd+1areaCmplx})
gives the finite part
\be
S^{ent}_{dS_4}
= {\pi l^2\over 2G_{4}}\,\Big({ia\over -\tau}\Big)\Big\vert_{-ia}
= {\pi l^2\over 2G_{4}}\,;\qquad
S^{ent}_{dS_5} = {\pi l^3\over G_5}\,\Big(-ia\,{\log(-1)\over 4a}\Big)
= {\pi^2 l^3\over 4G_5}\,.
\ee
These are half $dS_4$ and $dS_5$ entropy respectively\ (the $dS_5$
piece, using $\log(-1)=i\pi$, is similar to $dS_3$, arising from the
UV part, while the $dS_4$ piece arises from the IR $(-ia)$-end).
These are small subregions and there is no geometric hemisphere
interpretation here again: the $dS$ entropy piece arises nontrivially
from the complex $\tau$-plane path.  By comparison, the same $dS$
entropy piece arises in the IR surface area (maximal subregions) as
the Euclidean hemisphere contribution $S^{hemishere}_{dS}$ in
(\ref{dSd+1-StE}), along the Euclidean part $r=[0,l]$ of the time
contour $r=[0,l]\cup [l,R_c]$ in Figure~\ref{TimeContours}.

\section{Subregions from static patch to $\scri^+$, via light rays}
\label{sec:dSstatscri+}

Consider $dS_{d+1}$ in the static patch
\be\label{dSstatcoords}
ds^2 = -\Big(1-{r^2\over l^2}\Big) dt^2 + {dr^2\over 1-{r^2\over l^2}}
+ r^2 d\Omega_{d-1}^2\,.
\ee
The $r$-coordinate is an angular spatial coordinate in the static
patch $0\leq r<l$ (with $t$ being time), while $r$ is the time
coordinate in the future/past universe $l<r<\infty$ (while $t$ becomes
a spatial coordinate). See \cite{Narayan:2017xca,Narayan:2020nsc}
where timelike future-past extremal surfaces are analysed in detail in
this description.  See also \cite{Spradlin:2001pw} for various $dS$
coordinate systems.

\begin{figure}[h] 
\hspace{0.3pc}
  \includegraphics[width=18.6pc]{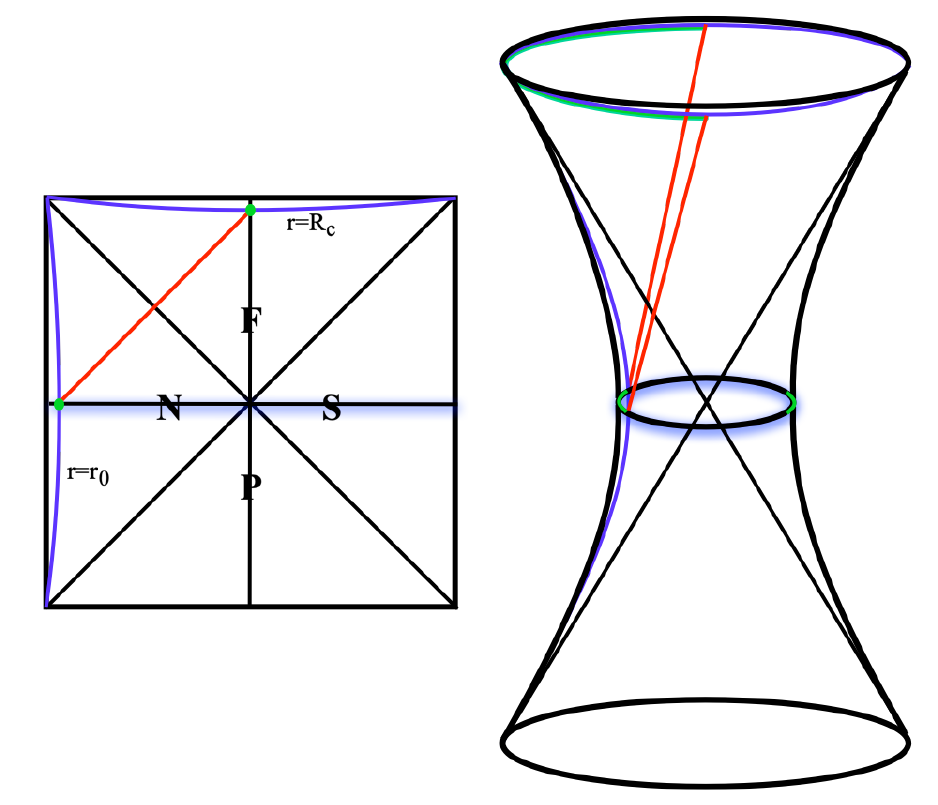} 
\hspace{2pc}
\begin{minipage}[b]{16.2pc}
\caption{{\label{dSlightrays}\!\! \footnotesize{Lorentzian $dS$ and
      lightrays in the $(t,r)$-plane Penrose diagram (left) from
      observers defining the $r=r_0$ cutoff slice on the $t=const$
      midslice in the static patch to the hemisphere subregion on the
      $t=0$ vertical slice at the cutoff future boundary at
      $r=R_c$. On the right, the horizontal circles represent the
      equatorial plane spheres, with sizes growing with $r$ from the
      midslice to $\scri^+$. A small subregion (green) inflates to a
      large one at $\scri^+$ along the lightrays (red). The blue
      shadow is the left-right extremal surface.} }}
\end{minipage}
\end{figure}
Observers in the static patch stationed at the North or South Poles
can send out future- or past-directed light rays to the future or
past boundary. From the $t=0$ constant time midslice, light rays
sent out intersect the (vertical) equatorial plane slice of the
sphere at the future boundary. If we instead regulate
$\scri^+$ imposing a cutoff, the corresponding equatorial plane
connects via lightrays to a cutoff timelike boundary on the $t=0$
midslice (see Figure~\ref{dSlightrays}). This gives a ``regulated'' or
``thickened'' observer at the North Pole. In more detail:\\
\noindent $\bullet$\ The spatial geometry on the $t=0$ slice around
the North Pole is the hemisphere
\be\label{dSstaticPatcht=0}
ds^2\Big\vert_{t=0} = {dr^2\over 1-{r^2\over l^2}}
+ r^2 d\Omega_{d-1}^2 = l^2 d\theta^2 + l^2\sin^2\theta d\Omega_{d-1}^2\,,
\quad r=l\sin\theta\,,\quad \theta=\big[0,{\pi\over 2}\big]\,.
\ee
(likewise we get a hemisphere in the static patch around the South Pole)\
Imposing a cutoff at $r=r_0$ gives the sphere $S^{d-1}$ on the $t=0$
slice of the timelike boundary $R_t\times S^{d-1}$. These $S^{d-1}$s are
latitude slices of the $S^d$ hemisphere.

\noindent $\bullet$\ Imposing a future boundary cutoff at $r=R_c\gg l$
gives a cylinder $R_t\times S^{d-1}$ in (\ref{dSstatcoords}). This
cylinder can be thought of as the sphere $S^d$ with metric\
$ds^2=\sin^2\theta({d\theta^2\over\sin^2\theta}+d\Omega_{d-1}^2)$\
after a conformal transformation removing the $\sin^2\theta$ factor.
The $t$-coordinate is spatial with range $[-\infty,\infty]$.
The $t=0$ slice is now a ``vertical'' slice passing through the middle
of the cylinder, which corresponds to an equatorial plane of the $S^d$.

From Figure~\ref{dSlightrays}, we see that a lightray from the North
Pole at $t=0$ (in {\bf $N$}) hits $\scri^+$ at the middle of {\bf $F$},
\ie\ $t=0$. The cutoff
surface $r=r_0$ is a timelike trajectory from the bottom left to
the top left corner. A lightray sent from $(t,r)=(0,r_0)$ reaches
$(t,r)=(0,R_c)$. However it is important to note that due to the
inflating property of $dS$, the subregion at $(0,R_c)$ on $\scri^+$
is a massively enlarged version of that on the cutoff timelike
boundary at $(0,r_0)$. To elaborate: lightray trajectories
in the Penrose diagram are described in the $(t,r)$-plane by
\be
ds^2=0\quad\ \Rightarrow\quad\
t_i-t_f=0-0=\int_{r_0}^{R_c} {dr\over |1-{r^2\over l^2}|}\,.
\ee
Here $t_i=0$ corresponds to the endpoint on the $t=0$ midslice in
{\bf $N$} ($r<l$) while $t_f=0$ is the endpoint on the $t=0$
vertical slice in {\bf $F$} ($r>l$). To evaluate this, we regulate
by symmetrically ``point-splitting'' the horizon as $r=l\pm \delta$
obtaining
\be
\int_{r_0}^{l-\delta} {dr\over 1-{r^2\over l^2}}\ +\
\int_{l+\delta}^{R_c} {dr\over {r^2\over l^2}-1}\ =\ 0
\ =\ {l\over 2} \log \Big({1+r/l\over 1-r/l}\Big)\Big\vert_{r_0}^{l-\delta}
\ +\ {l\over 2} \log \Big({r/l+1\over r/l-1}\Big)\Big\vert_{l+\delta}^{R_c}\ ,
\ee
\be
{\rm so}\quad r_0\ll l\,,\ R_c\gg l\ \ \Rightarrow\quad
\log {2l\over\delta} - {2r_0\over l} + {2l\over R_c}
- \log {2l\over\delta} = 0\ \ \Rightarrow\ \ r_0 \sim {l^2\over R_c}\ .
\ee
Stepping back, equating the integrands gives
\be\label{r<=l^2/r>}
{dr_<\over 1-{r_<^2\over l^2}} = -{dr_>\over 1-{r_>^2\over l^2}} \quad
\Rightarrow\quad r_< = {l^2\over r_>}\,.
\ee
This is essentially an inversion from the static patch $r_<$ to the
future universe $r_>$. So a small $r_<$-subregion maps to a large
$r_>$-subregion at $\scri^+$. In the extreme limit, the North Pole
(regulated as a small polar $S^d$ cap) maps to the maximal $S^d$
hemisphere subregion at $\scri^+$.\\
Likewise past-directed lighrays from the regulated South Pole hit
the the past boundary $\scri^-$ in the middle, \ie\ $(t,r)=(0,R_c)$.
So the South Pole regulated as a small polar $S^d$ cap maps to the
maximal $S^d$ hemisphere subregion at $\scri^-$, essentially using
(\ref{r<=l^2/r>}).

The relation (\ref{r<=l^2/r>}) holds for any $r_<$-subregion on the
$t=0$ midslice in the N/S static patches and their corresponding
subregions on the $t=0$ vertical slice in the future/past universes.
This suggests that the ``vertical'' future-past timelike extremal
surfaces connecting the future/past maximal subregions map to spatial
extremal surfaces connecting maximal subregions of the regulated North
and South Poles on the $t=0$ midslice in the North/South static
patches. We consider the $S^d$ on the $t=0$ constant time midslice in
the static patches and its boundaries on the $r=r_0$ cutoff surfaces
(which are $S^{d-1}$s): then we consider maximal (hemisphere)
subregions of these on the left and right static patches, and
extremal surfaces stretching between them. Since such surfaces
stretch from the boundary of the maximal (hemisphere) subregions of
the regulated Poles, they are codim-2 spacelike surfaces (roughly
rotating the timelike future-past extremal surfaces between
$\scri^\pm$). This left-right spatial extremal surface (horizontal
blue shadow in Figure~\ref{dSlightrays}), akin to the
Hartman-Maldacena surface in the $AdS$ black hole, wraps the
latitude $S^{d-2}$ and runs along the $r$-direction geodesically
from the boundaries of the maximal subregions of the regulated Poles
(green ball in $N$) to the horizon at $r=l$. Thus the area, using
(\ref{dSstaticPatcht=0}), is
\be\label{SLRdSd+1}
S_{LR} = 2\times {V_{S^{d-2}}\over 4G_{d+1}} \int_{r_0}^l {r_<^{d-2}\,dr_<
  \over \sqrt{1-{r_<^2\over l^2}}}\ \ \ \xrightarrow{\ r_0\ra 0\ }\ \ \
  {l^{d-1}\, V_{S^{d-1}}\over 4G_{d+1}}\,.
\ee
Thus we recover de Sitter entropy in the limit where the left/right
subregions approach the North/South Poles. The factor of 2 arises
from the left+right contributions. This blithe calculation can be seen
as such to be identical to twice the area contribution from the Euclidean
hemisphere \cite{Narayan:2022afv}, perhaps best interpreted
as ordinary (spatial) entanglement entropy between regulated $N,S$
static patch observer subregions: there is no holography here per se.
Let us evaluate this for $dS_3$ and $dS_4$, keeping $r_0$ nonzero but
small. We obtain
\bea\label{SLRdS3dS4}
&&  S_{LR} = 2\times {V_{S^0}\over 4G_3}\int_{r_0}^l
  {dr_<\over\sqrt{1-{r_<^2\over l^2}}}
  = {\pi l\over 2G_3} - {l\over G_3}\sin^{-1}{r_0\over l}\ \
  \xrightarrow{\ r_0\ll l\ }\ \ {\pi l\over 2G_3} - {r_0\over G_3}
  \quad [dS_3]\,,\qquad \nn\\
&&  S_{LR} = 2\times {V_{S^1}\over 4G_4}\int_{r_0}^l
{r_<\,dr_<\over\sqrt{1-{r_<^2\over l^2}}}
= {\pi l^2\over G_4} \sqrt{1-{r_0^2\over l^2}}\ \
\xrightarrow{\ r_0\ll l\ }\ \ {\pi l^2\over G_4} - {\pi r_0^2\over 2G_4}
\qquad\quad\ \ [dS_4]\,.
\eea
The leading terms are $dS_3$ and $dS_4$ entropy respectively,
while the subleading finite piece scales for small
$r_0$ as\ ${r_0\over G_3}\ [dS_3]$\ and ${r_0^2\over G_4}\ [dS_4]$.\

Note that there is no divergence in these areas: the sphere is finite.
The cutoff $r_0$ removes some of the region around the Poles in the
sphere $S^d$ and thereby induces a reduction in the leading entropy
term. This spatial extremal surface area (\ref{SLRdSd+1}) as de Sitter
entropy at leading order is reminiscent of
\cite{VanRaamsdonk:2009ar,VanRaamsdonk:2010pw}, with the bulk de
Sitter space and its entropy emerging via entanglement between the
left and right copies of the static patch. The anchored extremal
surfaces here are on slightly different footing however from the
discussions in \cite{Gupta:2025jlq}, \cite{Dong:2018cuv}, as well as
in \cite{Shaghoulian:2021cef}. Our analysis is coarse: it might be
interesting to explore this more carefully, perhaps building on
\cite{Coleman:2022lii}, and other $dS$ discussions
\eg\ \cite{Dong:2018cuv,Coleman:2022lii,Lewkowycz:2019xse,Coleman:2021nor,Shaghoulian:2021cef,Shaghoulian:2022fop,Cotler:2023xku,Franken:2023pni,Chang:2024voo,Chang:2025ays,Chakravarty:2025sbg}.


\section{Discussion}\label{sec:Disc}

As outlined in the Introduction, de Sitter extremal surfaces anchored
at $\scri^+$ have complex-valued areas best regarded as
pseudo-entropies
\cite{Doi:2022iyj,Narayan:2022afv,Doi:2023zaf,Narayan:2023ebn,Narayan:2023zen}.
We have seen that for large $\scri^+$ subregions (focussing on $dS_3$
for simplicity), no-boundary $dS$ timelike+Euclidean extremal surfaces
exist with transparent geometric interpretations (curves in the
spacetime/Penrose diagram) as do complex ones (which live solely in
some auxiliary $AdS$ space): their areas are identical\ (both families
can be understood as appropriate analytic continuations from $AdS$).
However, from (\ref{dS3tElimit}), the timelike+Euclidean surfaces stop
existing as the subregion size decreases (Figure~\ref{dS3surftE}):
only complex extremal surfaces exist for sufficiently small
subregions, analogous to Poincare $dS$ \cite{Narayan:2015vda}.  With
the extremal surface area integrals defined via time contours in the
complex time plane, we have found multiple extremal surfaces with
indistinguishable areas whose time contours can be deformed into each
other in the complex time plane without obstruction, that must be
regarded as equivalent for these pseudo-entropy purposes
(sec.~\ref{sec:dS3cmplxTimeC}). A simple example is the case of
maximal subregions in $dS_3$: contrast (\ref{dS3IR-tE}) for the
timelike+Euclidean area, and (\ref{dS3IR-cmplx}) for the imaginary
time path, both of which appear equivalent via deformations of their
time contours with fixed endpoints in the complex time plane
(Figure~\ref{TimeContours}). This also suggests
equivalences (sec.~\ref{sec:dS3cmplxGeom}) between a priori distinct-looking complex replica
geometries as a whole (possibly with foliations related by complex coordinate
transformations), the extremal surface arising from timelike+Euclidean
curves or complex horizons.  For small
subregions, we saw the $dS$ entropy term arising nontrivially along
complex time contours (with no hemisphere interpretation): similar
features arise in higher dimensional $dS_{d+1}$
(sec.~\ref{sec:dSd+1}). Generic subregions in higher dimensional $dS$
are difficult to analyse: it is likely that further features of
complex extremal surfaces will need to be understood here, towards
systematizing criteria for relevant extremal surface saddles, the
corresponding complex replica geometries and equivalences thereof.

It is worth noting however that for perturbations away from pure de
Sitter, extremal surface time contours will likely exhibit nontrivial
features. A related point is that the $-EAdS$ contours are likely to
make sense only when there exists a reasonable auxiliary $AdS$
interpretation for the geometries and the corresponding extremal
surfaces. This appears to not be true for the inflaton+metric
perturbations in no-boundary slow-roll inflation
\cite{Goswami:2024vfl} and the IR extremal surfaces there, which
appear best understood via the timelike+Euclidean contours (as
discussed at the end of sec.~\ref{sec:dS3cmplxTimeC}).

Analysing multiple small subregions in $dS_3$ reveals entropy
inequalities encoding $AdS$ ones via analytic continuation
(sec.~\ref{sec:dS3multsub}), consistent with previous discussions for
large subregions in \cite{Narayan:2023zen}. Subregion duality
(geometrically) is thus encoded in the $AdS$ continuations.  In recent
literature, similar features with complex extremal surfaces, generic
subregions and so on have been seen to arise in the study of timelike
entanglement: see
\eg\ \cite{Heller:2024whi,Milekhin:2025ycm,Guo:2025pru,Das:2023yyl,Nath:2024aqh,Katoch:2025bnh,Heller:2025kvp,Zhao:2025zgm,Fujiki:2025rtx,Li:2025tud,Guo:2025ase,Afrasiar:2025eam,Kanda:2026jyk,Hikida:2022ltr,Li:2022tsv,Jiang:2023ffu,Jiang:2023loq,Chu:2023zah,Chen:2023sry,Chen:2023eic,He:2023ubi,Guo:2024lrr,Fareghbal:2024lqa,Xu:2024yvf,Afrasiar:2024ldn,Harper:2025lav}.

We also discussed mapping future boundary subregions and those on
constant time slices in the static patch via lightrays in entirely
Lorentzian $dS$ (sec.~\ref{sec:dSstatscri+}).  Then, analogous to
future-past extremal surfaces between $\scri^\pm$, we exhibited
left-right extremal surfaces connecting small subregions in the $N/S$
static patches: their area equals $dS$ entropy in a limit. It would be
interesting to explore this further, possibly in relation to other
studies
\eg\ \cite{Dong:2018cuv,Coleman:2022lii,Lewkowycz:2019xse,Coleman:2021nor,
  Shaghoulian:2021cef,Shaghoulian:2022fop,Cotler:2023xku,
  Franken:2023pni,Chang:2024voo,Chang:2025ays,Chakravarty:2025sbg}.

\vspace{6mm}

{\footnotesize \noindent {\bf Acknowledgements:}\ \ It is pleasure to
  thank Wu-zhong Guo, Alok Laddha, Kanhu Nanda, Somnath Porey, Ronak
  Soni and Gopal Yadav for helpful conversations and comments on a
  draft.  This work is partially supported by a grant to CMI from the
  Infosys Foundation.  }

\vspace{-2mm}

{\footnotesize{
\bibliographystyle{JHEP} 
\bibliography{refsKN}
}}


\end{document}